# Towards an Artificial-Intelligence-Based Optical Scintillometer: Scaling Issue

G.A. Filimonov[1] and M.A. Vorontsov[1,2]

[1]Intelligent Optics Laboratory, Department of Electro-Optics and Photonics, School of Engineering, University of Dayton, 300 College Park, Dayton, Ohio 45469-2951 (gfilimonov1@udayton.edu)
[2]Optonica LLC, 2901 River End Court, Spring Valley OH 45370

*Abstract* —Atmospheric turbulence strength ($C_n^2$ parameter) sensing based on processing of intensity scintillation patterns with deep neural network (DNN) is considered. It is shown that DNN re-training with propagation distance change can be avoided by scaling of $C_n^2$ values obtained using a DNN trained for a nominal distance $L_0$. The required $C_n^2$ scaling factor can be obtained using either an analytical expression derived from the Kolmogorov turbulence theory (theory-based scaling), or through wave-optics numerical modeling and simulations (M&S-based scaling).

## I. INTRODUCTION

The refractive index structure parameter $C_n^2$ plays a special role in atmospheric optics signifying the importance of in-situ sensing of this turbulence strength characteristic [1,2]. Conventional electro-optical (EO) sensors used for $C_n^2$ measurements are based on the processing of large numbers of uncorrelated sensing data, which are needed for computation of a single $C_n^2$ value. Collection of these data takes a relatively long time (a few minutes). For this reason, these sensors cannot provide a sufficiently high temporal resolution, which is required for in-situ predictive performance assessment and forecasting of various atmospheric optics systems. Contrary to conventional $C_n^2$ sensing techniques, turbulence strength characterization based on the deep machine learning (ML) approach discussed here does not require collection of large datasets of uncorrelated measurements, but rather utilizes real-time analysis of sensing data features that are characteristic of a selected EO receiver, optical wave propagation geometry and turbulence condition [3,4]. This analysis is performed using a preliminary trained DNN-based signal processing system (DNN model). DNN model(s) training utilizes pre-recorded datasets composed of large numbers of synchronously acquired EO sensing data and $C_n^2$ measurements obtained side-by-side (along the same atmospheric propagation path) using a calibrated conventional $C_n^2$ sensor (e.g., scintillometer)[4]. Since DNN models are trained for a specified propagation path, they need to be re-trained with new datasets when the propagation distance is changed. Such datasets could be difficult or even possible to obtain, e.g., for $C_n^2$ sensing along the path to the moving target.

In this paper we show that re-training DNN models with propagation distance change can potentially be avoided via simple scaling of predicted $C_n^2$ values obtained using a "reference" DNN model trained for a nominal distance $L_0$. The required $C_n^2$ scaling factor can be obtained using either an analytical expression derived from the classical Kolmogorov turbulence theory (theory-based scaling), or through wave-optics numerical modeling and simulations (M&S-based scaling) of the EO sensor with the "reference" DNN over various propagation distances. The scaling concepts are illustrated here using numerical simulations of a ML-based EO sensor utilizing DNN-processing of short exposure intensity scintillation patterns for in-situ prediction of $C_n^2$ values [3,4].

## II. NUMERICAL MODELING AND SIMULATIONS (M&S) BASED SCALING

In the wave-optics based numerical simulations, we considered atmospheric propagation of a monochromatic (wavelength $\lambda$=1064 nm) collimated Gaussian laser beam (beacon beam) of diameter $d$=30 mm over a distance $L$ ranging from 0.9 km to 10 km. The turbulence-induced refractive index random fluctuations were assumed to be uniform along the propagation path and obey the Kolmogorov refractive index power spectrum model [1]. To simulate refractive index structure parameter diurnal variation with a simple model, the following dependence of $C_n^2(m)$ (ground truth curve in Fig. 1) was used [3]:

$$C_n^2(m) = C_{n,0}^2 \left[\cos(3\pi m / M_{SIM}) + 1\right] + b, \qquad (1)$$

where $m = 1,\ldots, M_{SIM}$ is an index of instances in the simulation datasets, and $C_{n,0}^2 = 2.0 \cdot 10^{-14}\,\text{m}^{-2/3}$ and $b = 1.0 \cdot 10^{-17}\,\text{m}^{-2/3}$ are parameters selected to match the typical range of $C_n^2$ change. The $C_n^2(m)$ dependence described by Eqn. (1) is shown in Fig. 1 (a) (ground truth curve). For each $C_n^2(m)$ value the turbulence-induced refractive index variations along the propagation path were modeled by a set of ten thin statistically independent random phase screens that were equidistantly distributed along the propagation path. The intensity distribution at the optical receiver plane $I(\mathbf{r}, m)$ (scintillation pattern) was obtained by simulating propagation of the beacon beam over the $m$th turbulence realization corresponding to the defined by Eqn (1) $C_n^2(m)$ value using the split-step operator technique [5]. For $C_n^2$ prediction based on processing of laser beam intensity scintillation patterns the DNN model (Cn^2Net model) described in [3] was utilized. The Cn^2Net model was trained using a dataset composed of $M_{SIM}$ =15,000 (15K) instances comprised of $\{C_n^2(m)\}$ values and corresponding scintillation frames $\{I_{ref}(\mathbf{r}, m)\}$ computed for the propagation distance $L_0$ = 7 km (reference dataset). Additional datasets (inference datasets) with equal numbers of instances ($M_{SIM}$ =15K) were generated for different propagation distances. The "reference" DNN model pre-trained for $L_0$ =7 km was utilized in a set of ML experiments with inference datasets (using scintillation frames $\{I_L(\mathbf{r}, m)\}$ computed for different $L$ as the DNN inputs). The corresponding DNN output values $\{C_n^2(m,L)\}$ obtained in inference experiments for $L$=3 km, 7 km and $L$=10 km are shown in Fig. 1(a) by colored dots. As expected, the smallest deviations $\Delta(m,L) = C_n^2(m,L) - C_n^2(m)$ between the DNN output $C_n^2(m,L)$ (prediction) and the true refractive index value $C_n^2(m)$ were observed for the inference dataset corresponding to $L_0 = 7$km [green dots in Fig. 1(a)]. With a change in the propagation distance $L$, the prediction errors $\Delta(m,L)$ increased as indicated in Fig. 1(a) by the black dots for $L$=3 km and blue dots for $L$=10 km. The mean square prediction error (MSE) averaged across the entire $C_n^2$ range, defined as $\bar{\varepsilon}(L) = \sum_{m=1}^{M_{SIM}} \Delta^2(m,L)$, is shown by the black curve in Fig. 1(b) as a function of the propagation distance $L$. The MSE prediction error $\bar{\varepsilon}(L)$ has a minimal value [$\bar{\varepsilon}(L_0) \approx 0.07$] for the dataset corresponding to the intensity scintillation patterns obtained for the nominal distance $L = L_0$ used for the DNN model training. The error $\bar{\varepsilon}(L)$ increases by a factor of 4.6 for the shortest ($L = 0.125L_0$) and by factor 1.8 for the longest ($L = 1.4L_0$) distances. The $C_n^2$ prediction error can be corrected by multiplying the DNN output value $C_n^2(m,L)$ by a scaling factor $\gamma(L)$ that depends on propagation distance $L$. It is assumed that this distance is either known a priori or can be measured in-situ.

For each fixed propagation distance $L$ the scaling factor can be found via minimization of the MSE for "corrected" deviations $\Delta_\gamma(m,L) = \gamma(L)C_n^2(m,L) - C_n^2(m)$ between the scaled by factor $\gamma(L)$ DNN output $C_n^2(m,L)$ and the true refractive index value $C_n^2(m)$ : $\gamma_{opt}(L) = \arg\min_{\gamma(L)} \sum_{m=1}^{M_{SIM}} [\gamma(L)C_n^2(m,L) - C_n^2(m)]^2$. In the simulations the optimal (for each propagation distance $L$) scaling factor $\gamma_{opt}(L)$ was computed and used for correction of the DNN output values $C_n^2(m,L)$. The products $\{\gamma_{opt}(L) \cdot C_n^2(m,L)\}$ were considered as corrected $C_n^2$ predictions. The MSE $\bar{\varepsilon}_\gamma(L) = \sum_{m=1}^{M_{SIM}} [\gamma_{opt}(L)C_n^2(m,L) - C_n^2(m)]^2$ corresponding to the corrected predictions is presented in Fig. 1 (b) by the blue curve. The results demonstrate significant improvement in accuracy of $C_n^2$ prediction (compare the MSE curves obtained without and with M&S-based scaling).

### III. SCALING BASED ON AN ANALYTICAL EXPRESSION FOR THE FRIED PARAMETER (THEORY-BASED SCALING)

The characteristic size $l_s$ of turbulence-induced speckles in scintillation patterns represents a distinguishing spatial feature that is "captured" by the DNN during the training process. In a similar way, a characteristic correlation length for turbulence-induced phase aberrations, as described by the Fried parameter $r_0$, can be considered as a spatial feature that is analogous to $l_s$. This suggests that the analytical expression describing $r_0$ dependence on the propagation distance $L$ in the framework of Kolmogorov turbulence theory can be utilized for scaling the DNN model output values. For a spherical wave the Fried parameter is given by the following expression [2]:

$$r_0 = \left(0.423 C_n^2 k^2 L\right)^{-3/5}, \quad (2)$$

where $k = 2\pi/\lambda$ and $\lambda$ is wavelength. If the assumption above about the similarity of the spatial feature parameters $l_s$ and $r_0$ is correct, the "theory-based" scaling factor $\gamma_T(L)$ can be obtained from Eqn. (2) by equating Fried parameter values associated with the reference DNN and the corresponding $r_0$ values computed based on scaled $C_n^2$ predictions for an arbitrary distance $L$: $r_0 = \left(0.423 C_n^2 k^2 L_0\right)^{-3/5} = \left(0.423 \gamma_T(L) C_n^2 k^2 L\right)^{-3/5}$. Correspondingly, for the scaling factor $\gamma_T(L)$ we obtain: $\gamma_T(L) = L_0/L$. The MSE $\bar{\varepsilon}_F(L) = \sum_{m=1}^{M_{SIM}}[\gamma_F(L)C_n^2(m,L) - C_n^2(m)]^2$ corresponding to the corrected predictions is presented in Fig. 1(b) by the red curve. Surprisingly, the scaling factor derived from the classical expression for the Fried parameter (theory-based scaling) provides nearly identical $C_n^2$ prediction accuracy as more computationally expensive MSE-based scaling (compare curves obtained with M&S- and theory-based scaling) in Fig. 1(b).

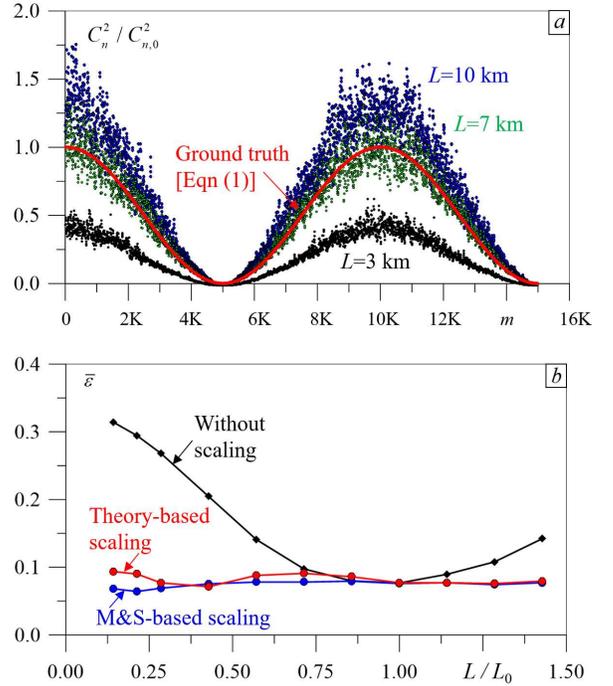

Fig. 1: (a) Normalized DNN output $C_n^2(m,L)$ values (dots) obtained in machine learning experiments with inference datasets for $L$=3 km (black dots), $L$=7 km (green dots) and $L$=10 km (blue dots) along the truth $C_n^2(m)$ values (solid curve); (b) averaged prediction error vs propagation distance $L/L_0$ for the "reference" DNN pre-trained for $L/L_0 = 1$ without predicted $C_n^2$ values scaling and with M&S- and theory-based scaling. Normalization factors are $L_0$=7 km and $C_{n,0}^2 = 10^{-14}$ m$^{-2/3}$.

### REFERENCES
[1] A.N. Kolmogorov, *Turbulence: Classic Papers on Statistical Theory*; Interscience: New York, NY, USA, 1961.
[2] V.I. Tatarskii, *Wave Propagation in a Turbulent Medium*; McGraw-Hill: New York, NY, USA, 1961.
[3] A.M. Vorontsov, M.A. Vorontsov, G.A. Filimonov, and E. Polnau, *Applied Science*, vol. 10, p. 8136, 2020.
[4] D.L.N. Hettiarachchi, E. Polnau, and M.A. Vorontsov, in *Free-Space Laser Communications XXXIV*. SPIE, 2022, vol. 11993, pp. 288-291.
[5] J.A. Fleck, J.R. Morris, and M.D. Feit, *Appl. Phys.* vol.10, p. 129, 1976